\documentclass[conference]{IEEEtran}
\IEEEoverridecommandlockouts
\usepackage{cite}
\usepackage{listings}
\usepackage{color}
\usepackage{placeins}
\usepackage{afterpage}
\usepackage{amsmath,amssymb,amsfonts}
\usepackage{algorithmic}
\usepackage{graphicx}
\usepackage{textcomp}
\usepackage{longtable}
\usepackage{array}
\usepackage{graphicx,color}
\usepackage{float}
\usepackage{enumerate}
\usepackage{makecell}
\usepackage{longtable}
\usepackage{graphicx}
\usepackage{lipsum}
\usepackage{float}
\usepackage{multicol}
\usepackage{multirow}
\usepackage{xcolor}
\def\BibTeX{{\rm B\kern-.05em{\sc i\kern-.025em b}\kern-.08em
    T\kern-.1667em\lower.7ex\hbox{E}\kern-.125emX}}
\begin{document}
\title{SDN-Based Smart Cyber Switching (SCS) for Cyber Restoration of a Digital Substation\\

}

\author{
\IEEEauthorblockN{
Mansi Girdhar\IEEEauthorrefmark{1},
Kuchan Park\IEEEauthorrefmark{1},
Wencong Su\IEEEauthorrefmark{1},
Junho Hong\IEEEauthorrefmark{1},
Akila Herath\IEEEauthorrefmark{2},
Chen-Ching Liu\IEEEauthorrefmark{2}
}
\IEEEauthorblockA{
\textit{Department of Electrical and Computer Engineering}
}
\IEEEauthorblockA{
\IEEEauthorrefmark{1}~\textit{University of Michigan-Dearborn, USA} \\
\IEEEauthorrefmark{2}~\textit{Virginia Tech, Blacksburg, VA, USA}
}
\IEEEauthorblockA{
gmansi@umich.edu, kuchan@umich.edu, wencong@umich.edu, jhwr@umich.edu,
akilaasansana@vt.edu, ccliu@vt.edu
}
}

\maketitle
\pagestyle{plain}

\begin{abstract}
In recent years, critical infrastructure and power grids have increasingly been targets of cyber-attacks, causing widespread and extended blackouts. Digital substations are particularly vulnerable to such cyber incursions, jeopardizing grid stability. This paper addresses these risks by proposing a cybersecurity framework that leverages software-defined networking (SDN) to bolster the resilience of substations based on the IEC-61850 standard. The research introduces a strategy involving smart cyber switching (SCS) for mitigation and concurrent intelligent electronic device (CIED) for restoration, ensuring ongoing operational integrity and cybersecurity within a substation. The SCS framework improves the physical network's behavior (i.e., leveraging commercial SDN capabilities) by incorporating an adaptive port controller (APC) module for dynamic port management and an intrusion detection system (IDS) to detect and counteract malicious IEC-61850-based sampled value (SV) and generic object-oriented system event (GOOSE) messages within the substation's communication network. The framework's effectiveness is validated through comprehensive simulations and a hardware-in-the-loop (HIL) testbed, demonstrating its ability to sustain substation operations during cyber-attacks and significantly improve the overall resilience of the power grid.
\end{abstract}


\section{Introduction}
The power system is a complex network comprising generation, transmission, and distribution stages, where substations play a pivotal role in altering voltage levels. Traditionally, substations have relied on hardwired communication lines to manually operate devices such as switches and circuit breakers (CBs), resulting in cumbersome and maintenance-heavy configurations. In addition, legacy communication protocols (e.g., DNP3 and Modbus) exacerbate complexities due to inconsistent data mapping across different vendors' products \cite{9369743}.

The International Electrotechnical Commission (IEC) introduced the IEC 61850 standard to mitigate these challenges, offering substantial benefits such as multi-vendor interoperability, reduced configuration efforts, cost-effective installation, and high-speed, Ethernet-based communication for time critical signals. While these advancements have transformed substations into more efficient and intelligent systems, they have also introduced new cybersecurity vulnerabilities. The digitization and increased use of ICT in substations have exposed these critical infrastructures to risks of cyber-attacks, potentially leading to system failures and significant operational disruptions.

Recent cyber-attacks on critical infrastructure, such as the coordinated attack on Ukraine's power grid that disabled 30 substations and left approximately 230,000 residents without electricity for six hours, underscore the urgent need for improved cybersecurity measures. Despite numerous efforts to develop cybersecurity standards and defense mechanisms for power grids, significant limitations remain, including: 

\begin{itemize}
    \item Limited Comprehensive Cybersecurity Solutions: Many approaches address isolated aspects of cybersecurity, lacking a holistic approach to mitigate all identified vulnerabilities.
    \item Challenges in Localization and Isolation of Attacks: Current methods often emphasize intrusion detection without effective mechanisms to localize and isolate attacks within substations. The previous work \cite{10688802} by the same authors focused on the localization of malicious hosts.
    \item Inadequate Real-Time Response Capabilities: Existing solutions generally lack the dynamic reconfiguration abilities to respond effectively to real-time cyber threats.
\end{itemize}

Software-defined networking (SDN) is an emerging paradigm known for its dynamic, controllable, and flexible nature, suitable for high-bandwidth and dynamic applications. It separates the control and forwarding planes, thus enhancing network scalability and flexibility. However, this separation also increases the attack surface, necessitating robust security measures. Centralized SDN management is susceptible to various cyber-attacks, such as fault injection and distributed denial-of-service (DDoS) attacks. In addition, some current commercial SDN switches lack the functionality for remote port reconfiguration. This limitation hampers the ability to isolate compromised network segments effectively. Other commercial SDN switches are prohibitively expensive, hindering their widespread adoption in substation cybersecurity applications. As a result, malicious packets can continue to spread during cyber-attacks, potentially disrupting substation operations.

Redundant protective intelligent electronic devices (PIEDs) are vital for continuous operation in digital substations, as attackers can compromise PIED functions by injecting malicious sampled values (SV) or generic object-oriented system events (GOOSE) packets. While previous research has focused on creating backup IEDs \cite{LIM2016151}, these efforts have not fully addressed the cybersecurity.

To address these challenges, this paper introduces an innovative smart cyber switching (SCS) framework that incorporates concurrent intelligent electronic device (CIED) integration, representing the first implementation of these methods in tandem. This approach leverages the strengths of both SCS and CIED technologies, providing an advanced, integrated solution for real-time, resilient cyber-physical interactions within digital substations. The SCS framework comprises:

\begin{itemize}
    \item Adaptive Port Controller (APC): A module that dynamically manages OpenFlow table rules and policies to reconfigure the network in real-time.
    \item Network-Based Intrusion Detection System (IDS): An advanced IDS \cite{6786500} designed to detect severe cyber-attacks, triggering the SCS to isolate compromised devices and invoking CIED to take over essential protection functions.
\end{itemize}

The primary contributions of this paper include:

\begin{enumerate}
    \item Development and implementation of the SCS framework utilizing APC for real-time, dynamic network reconfiguration to isolate cyber threats.
    \item Introduction of CIED that replicates the protection functions of compromised physical IEDs, ensuring continuous protection and control.
    \item Demonstrations of the framework’s effectiveness in mitigating SV and GOOSE cyber-attacks, ensuring resilient substation operations.
\end{enumerate}

This integrated approach offers a comprehensive solution to the cybersecurity challenges faced by modern substations, enhancing resilience and reliability.

The remainder of the paper is structured as follows: Section II provides the information of the existing hardware-in-the-loop (HIL) testbed. Section III elaborates on the proposed SCS framework and its components. Section IV focuses on the application of the proactive substation security algorithm (PSSA) for attack mitigation and isolation using SCS. Section V presents simulation and validation results and describes the transition of protection functions to CIED. Section VI concludes with recommendations for future work.

\section{Hardware-In-The-Loop (HIL) Testbed}
\begin{figure}[htb!]
\centering
\includegraphics[width= 0.5\textwidth, height = 2.45in]{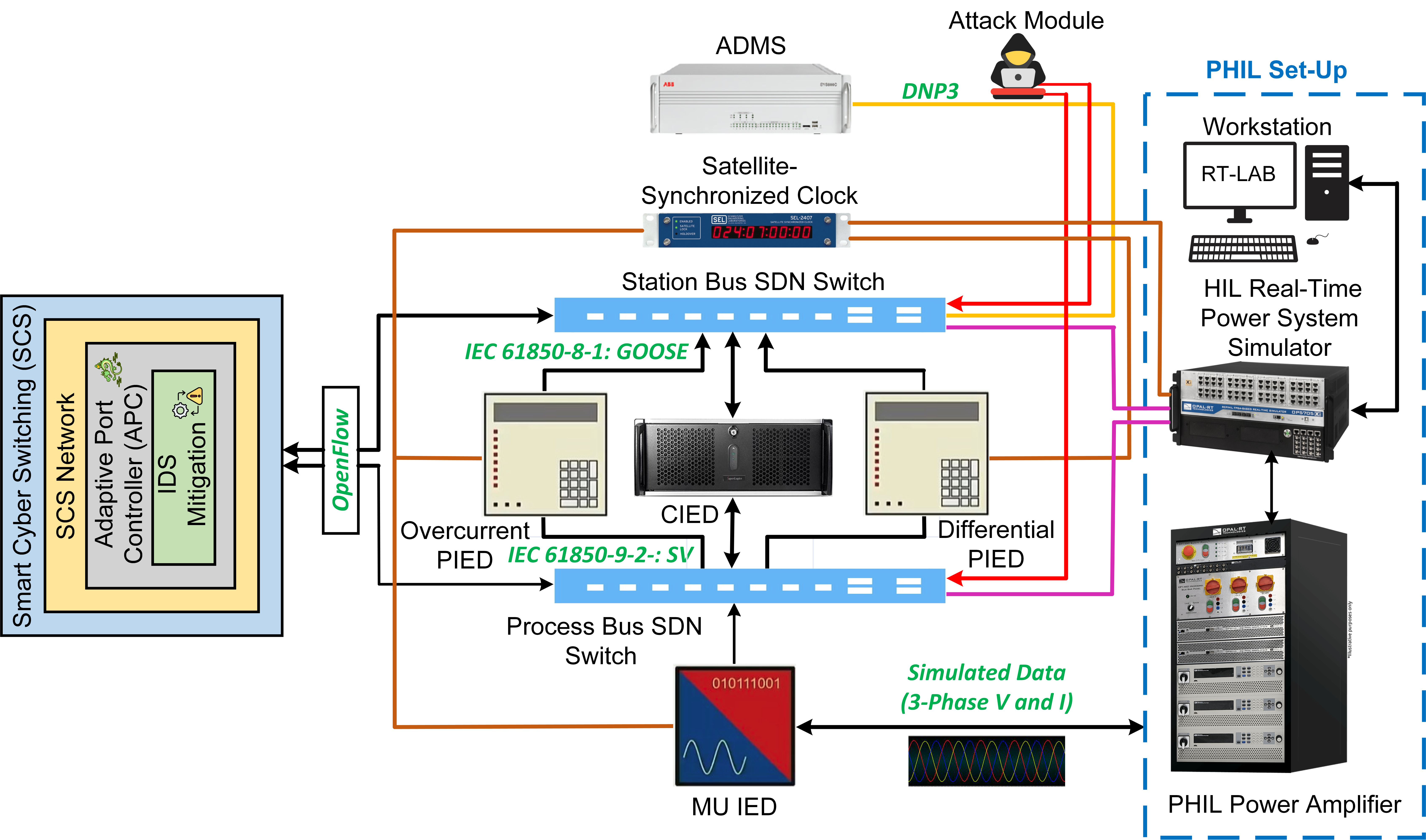}
\caption{HIL testbed configuration for evaluating substation cybersecurity.}
\label{fig:PESGM_01}
\end{figure}

This section provides a high-level overview of the HIL testbed, illustrated in Fig.~\ref{fig:PESGM_01}, before detailing the proposed SCS environment. The HIL testbed is essential to the cybersecurity architecture for smart grids, simulating a substation’s environment with a mix of software-defined and real-time physical devices \cite{10578257}. It includes key components such as a satellite-synchronized clock for precise synchronization, a commercial relay functioning as a merging unit (MU), overcurrent and differential PIEDs for monitoring and control, a real-time power system simulator and amplifier for simulations and prototyping, and SDN switches for enhanced network flexibility and control. These components ensure efficient power management using IEC 61850-based protocols such as SV and GOOSE. The setup involves a complex network where the power simulator, IEDs, SDN switches, and SDN controller collaborate to secure the system. For instance, the power simulator sends digitized signals via MU, and then sends SVs to overcurrent PIED using the SDN process bus switch. If an overcurrent fault is detected, PIED sends GOOSE messages through the SDN station bus switch to trip its associated CB. The testbed simulates conditions from normal operations to cyber-attacks, where undetected SV or successful GOOSE cyber-attacks could illicitly trigger CB trips, emphasizing the need for robust cybersecurity measures.

\vspace{-1.25 pt}

\section{Proposed Smart Cyber Switching (SCS) Framework}

\begin{figure}[htb!]
\centering
\includegraphics[width= 0.5\textwidth, height = 1.9in]{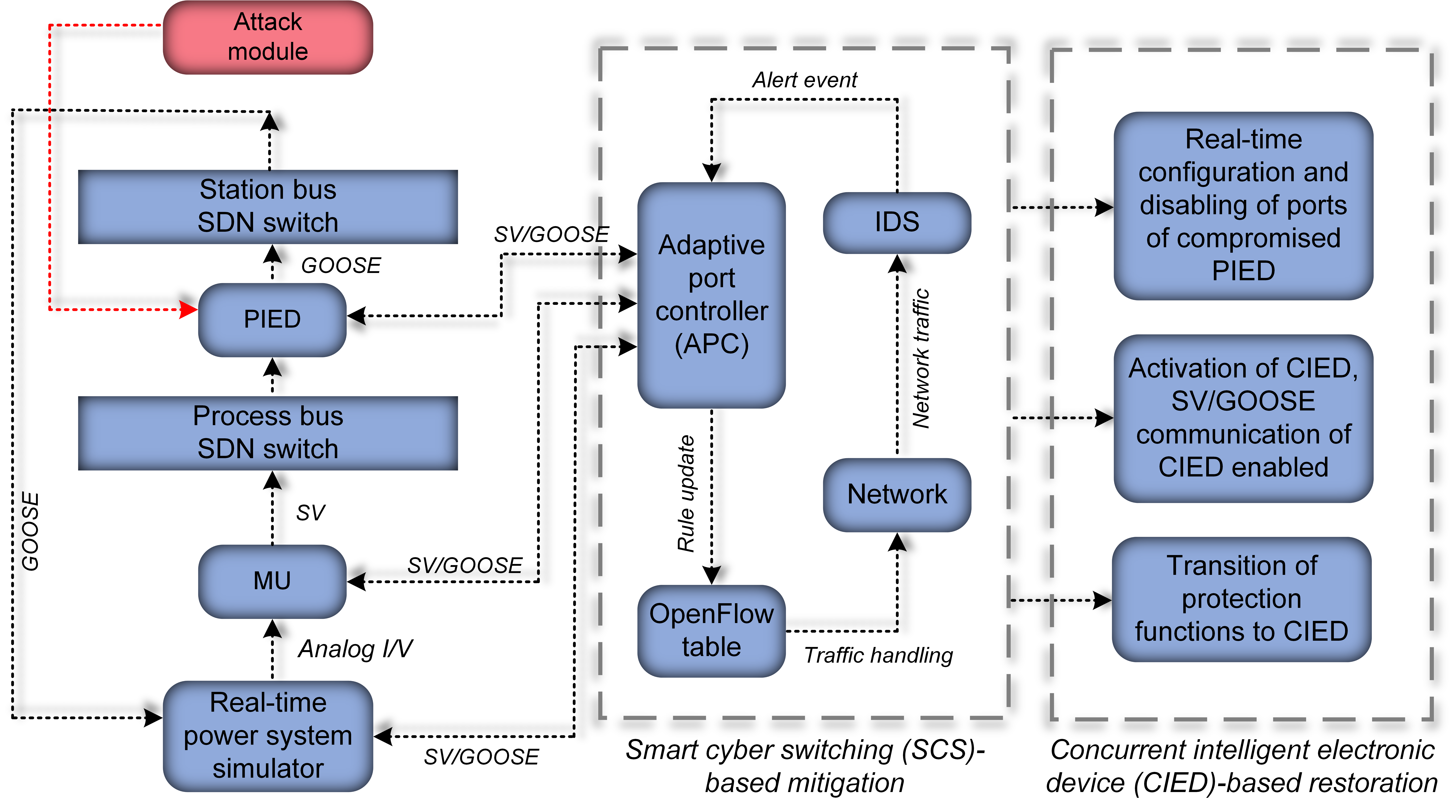}
\caption{Proposed smart cyber switching (SCS)-based mitigation using adaptive port controller (APC) and restoration using concurrent IED (CIED).}
\label{fig:Design_architecture}
\end{figure}
\vspace{-1.25 pt}

The design architecture of the proposed SCS framework is illustrated in Fig.~\ref{fig:Design_architecture}. This SCS network topology supports IEDs, transmitting IEC 61850-based SV and GOOSE packets, and employing IDS for network security. The proposed SCS network can be scaled to accommodate more hosts or larger substations by updating the topology script. The SCS framework leverages SDN's flexibility and programmability to create a dynamic network capable of detecting, mitigating, and isolating cyber threats efficiently. Centralizing network control allows for real-time reconfiguration, enhancing the resilience of substation automation systems against sophisticated cyber-attacks.

The SCS network is implemented on an Ubuntu 22.04 host using Mininet, an open-source network emulator, and Ryu SDN controller \cite{10318551}. The libIEC61850 library ensures proper handling of SV and GOOSE communications, while Scapy is used for custom packet parsing and building. The SCS APC dynamically manages network ports, optimizing utilization and traffic flow. It includes several components: an adaptive controller for real-time decision-making, a port management system to enforce these decisions, monitoring and analytics for feedback, and a communication interface for seamless interactions between controller and network devices.

The SCS APC automates port reconfiguration, optimizing network performance and adapting to traffic changes. It ensures efficient resource allocation, scales easily, and simplifies network management through centralized control. This environment provides a robust platform for developing advanced cybersecurity measures, enabling smart grids to counter sophisticated cyber threats. As shown in Fig.~\ref{fig:Design_architecture}, the SCS framework processes data from MU, PIEDs, simulators, and SDN switches. The data, including SVs and GOOSE messages, is analyzed to detect anomalies such as feeder and transformer faults. The SCS-IDS monitors network traffic for threats, updating OpenFlow table rules as necessary. Upon detecting anomalies, the IDS alerts the SCS APC, which then updates network routing rules. The SDN switch enforces these changes, directing or blocking traffic as needed. By tracking traffic changes, the SCS framework identifies faults and isolates malicious hosts. The SCS APC then handles mitigation and CIED-based restoration. Supporting multiple OpenFlow versions, the SCS APC allows real-time adjustments, effectively blocking malicious traffic and isolating compromised devices to prevent cyber-attacks. The integration of CIED significantly enhances the resilience and reliability of substation automation systems. CIED emulates the protection functions of physical IEDs, such as overcurrent and differential PIEDs, adding redundancy and security. Their primary goal is to maintain continuous protection and control, even when physical IEDs are compromised by cyber-attacks. By transitioning critical functions to software-based IEDs, the CIED framework ensures uninterrupted substation operations, preserving grid integrity. CIED utilizes substation configuration language (SCL) engineering, accesses all operational and protection data, and runs continuously without manual activation, minimizing unprotected time.

\begin{figure*}[htb!]
\centering
\includegraphics[width= 1.0 \textwidth, height =2.25in]{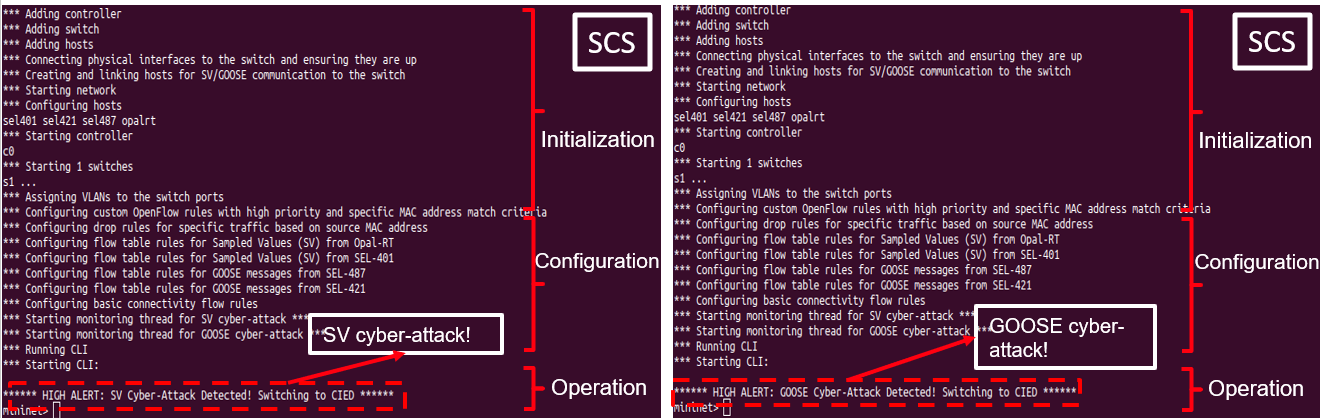}
\caption{SCS network initialization and configuration log with real-time SV and GOOSE cyber-attacks detection and response.}
\label{fig:Wireshark_attack}
\end{figure*}

\section{Proactive Substation Security Algorithm (PSSA)}
The objective of the proposed PSSA is to minimize the overall disruption caused by cyber-attacks while ensuring the functionality and reliability of the substation network.

Let \textit{\textbf{P}} is the set of \textit{\textbf{N}} PIEDs,
\begin{equation}
    P=\left\{ P_{1},P_{2},...,P_{N} \right\},
\end{equation} where \textit{\textbf{P}}$_{i}$ denotes each PIED and its corresponding weight, representing the impact of each PIED in the network.

Further, attack detection status is defined as: \begin{equation}A_{i}\epsilon\left\{0,1 \right\}\end{equation} for each PIED, where 0 means no cyber-attack on \textit{\textbf{P}}$_{i}$ and 1 means occurrence of cyber-attack.

Also, attack status vector \textit{\textbf{A}} represents the attack status of all \textit{\textbf{N}} PIEDs such that,
\begin{equation}
    A=\left\{A_{1},A_{2},...,A_{N} \right\}.
\end{equation}

There are a few decision variables defined, e.g.,
\textit{\textbf{D}}$_{i}$ is a binary variable that determines whether a PIED is attacked or not. Ideally,
\begin{equation}
    D_{i}= A_{i}, \forall i \in \left\{ 1,2,...,N \right\}.
\end{equation}
 It means if \textit{\textbf{P}}$_{i}$ is attacked \textit{\textbf{({A}$_{i}$ = 1)}}, it must be disabled \textit{\textbf{({D}$_{i}$ = 1)}} and if \textit{\textbf{P}}$_{i}$ is not attacked \textit{\textbf{({A}$_{i}$ = 0)}}, it remains enabled \textit{\textbf{({D}$_{i}$ = 0)}}. However, there could be scenarios where, due to operational constraints, a PIED cannot be immediately or effectively disabled even if an attack is detected, leading to, \begin{equation}
    D_{i} \neq A_{i}, \forall i \in \left\{ 1,2,...,N \right\}.
\end{equation}

Similarly, \textit{\textbf{E}} is a binary variable that determines if the CIED is enabled. To instantiate, CIED is enabled \textit{\textbf{({E} = 1)}} if any (at least one PIED) \textit{\textbf{P}}$_{i}$ is under attack. 

\begin{equation}
    E=min(1, \sum_{i=1}^{N}A_{i}).
\end{equation}

Another binary variable, \textit{\textbf{F}}$_{i}$, indicates whether a flow rule update is required to redirect traffic to the CIED when \textit{\textbf{P}}$_{i}$ is under attack  \textit{\textbf{({F}$_{i}$ = 1)}} or not \textit{\textbf{({F}$_{i}$ = 0)}}.

The objective function is formulated as follows:

\begin{equation}
    minimize\left[\sum_{i=1}^{n}(P_{i}\cdot D_{i})+\gamma E +\sum_{i=1}^{n}(A_{i}\cdot F_{i})\right].
\end{equation}

The parameter \textbf{$\gamma$} is defined as the weight associated with enabling the CIED. It quantifies the relative importance of enabling the CIED. When defining \textbf{$\gamma$}, several factors may be considered, including: a balance between redundancy, reliability, and resource cost or historical data or simulations providing insights into the impact of enabling the CIED.

\section{Simulation and Validation}
Simulation and validation are critical to ensuring that the SCS framework operates as intended for both normal operations and cyber-attack scenarios, ensuring that the defensive mechanisms are effective and reliable.

\subsubsection{Cyber-Attack Scenarios}
Simulations of SV and GOOSE cyber-attacks demonstrate the SCS framework's capability to isolate compromised devices upon detecting malicious activities. SCS APC’s dynamic flow rule adjustments isolate the attacked devices and transit protection functions of the compromised PIEDs to CIED seamlessly as a restoration measure, hence ensuring network stability. The SCS interface outputs, as shown in Fig.~\ref{fig:Wireshark_attack}, indicate detected SV and GOOSE attacks and corresponding actions, specifically highlighting the transition to CIED. 


\begin{figure}[htb!]
\centering
\includegraphics[width= 0.45\textwidth, height =2.5in]{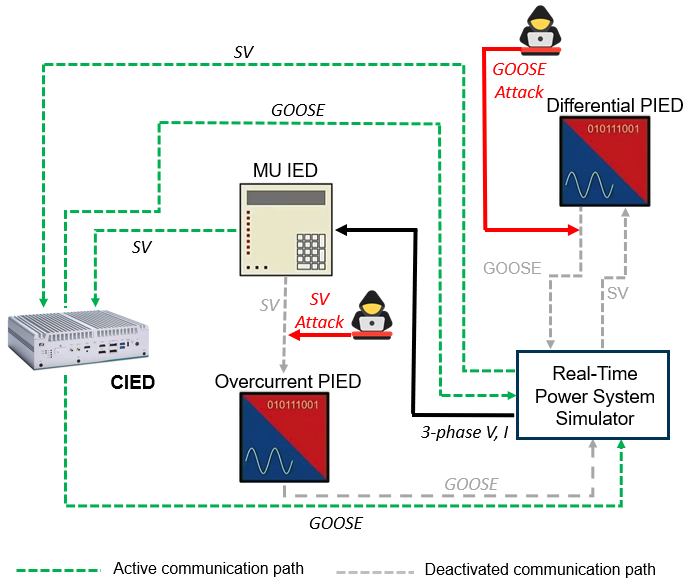}
\caption{Transition of protection functions of overcurrent PIED and differential PIED to CIED on the occurrence of cyber-attacks.}
\label{fig:CIED_new}
\end{figure}

\begin{figure}[htb!]
\centering
\includegraphics[width= 0.5\textwidth, height = 2.5in]{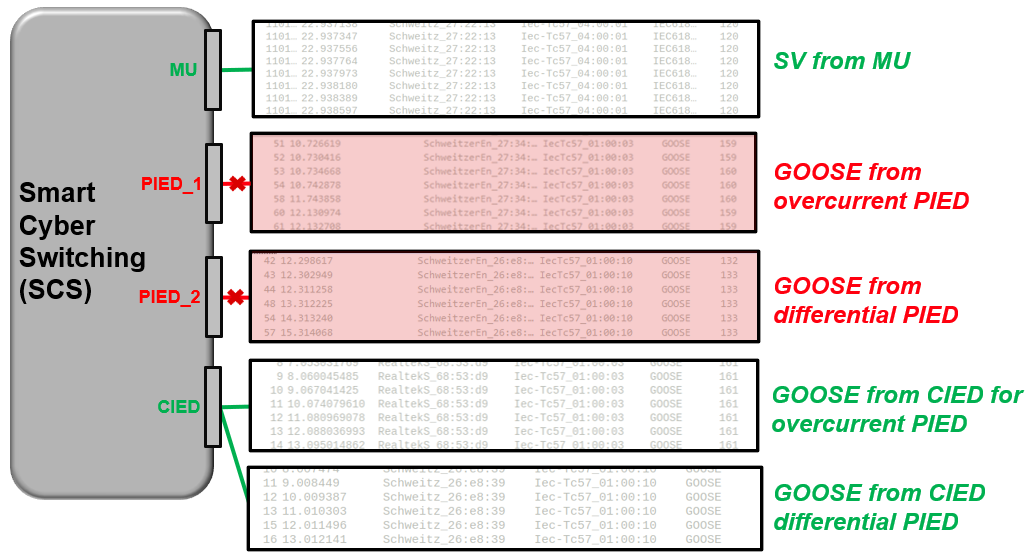}
\caption{GOOSE packets from CIED that performs the protection functions of overcurrent and differential PIEDs when cyber-attacks occur on the digital substation.}
\label{fig:Wireshark_new}
\end{figure}

To illustrate the efficacy of the CIED framework, consider the following case study involving SV and GOOSE attack scenarios. Fig.~\ref{fig:CIED_new} demonstrates the transition of protection functions from overcurrent and differential PIEDs to CIED when there are SV and GOOSE cyber-attacks, respectively. It shows how the communication ports corresponding to the overcurrent PIED and differential PIED are disabled, and corresponding ports of CIED are enabled on the application of the SCS framework.

\subsubsection{SV Attack Scenario}

\begin{itemize}
    \item Attack Detection: An SV cyber-attack is detected by the SCS IDS, compromising the overcurrent PIED.
    \item Port Blocking: The compromised ports are dynamically disabled to prevent further propagation of the attack.
    \item CIED Activation: The SCS framework automatically activates the CIED to take over the protection functions of the overcurrent PIED, ensuring continuous operation.
    
\end{itemize}

\lstset{frame=tb,
  language=Python,
  aboveskip=3mm,
  belowskip=3mm,
  showstringspaces=false,
  columns=flexible,
  basicstyle={\small\ttfamily},
  numbers=none,
  numberstyle=\tiny\color{black},
  keywordstyle=\color{blue},
  commentstyle=\color{dkgreen},
  stringstyle=\color{mauve},
  breaklines=true,
  breakatwhitespace=true,
  tabsize=3
}
\begin{lstlisting}
****** HIGH ALERT: SV Cyber-Attack Detected! Switching to CIED *****
\end{lstlisting}

Fig.~\ref{fig:Wireshark_new} clearly shows the Wireshark capture of GOOSE packets by the CIED when there is an SV cyber-attack, due to which CIED takes over the protection function of the overcurrent PIED as it gets compromised.

\paragraph{Simulation Setup and Results}
In the simulation setup of the current HIL testbed, two PIEDs are considered and the following assumptions are made regarding the weights of each PIED and CIED. Number of PIEDs: \textit{\textbf{N}} = 2, Weight of PIEDs: \textit{\textbf{P}} = [10,15], 
Weight for enabling the CIED: \textbf{$\gamma$} = 5.

When there are no cyber-attacks on the digital substation, the outcome of the optimal objective value is 0, as D = [0,0], E = 0, and F = [0,0]. This signifies that  no PIEDs are attacked; therefore, no actions are required and hence, all PIEDs remain operational, and CIED is not enabled by the SCS. However, when there is an SV attack, optimal objective value becomes 16, as D = [1,0], E = 1, and F = [1,0]. This signifies that PIED 1 (i.e., overcurrent PIED) is attacked and disabled. Hence, CIED is enabled to take over, and flow rules are updated accordingly.

\subsubsection{GOOSE Attack Scenario}

\begin{itemize}
    \item Attack Detection: A GOOSE cyber-attack is detected, compromising the differential PIED.
    \item Port Blocking: Similar to the previous case, compromised network segments are isolated, maintaining the integrity of the rest of the network.
    \item CIED Activation: SCS module activates the corresponding CIED that performs differential protection functions.
    
\end{itemize}

\lstset{frame=tb,
  language=Python,
  aboveskip=3mm,
  belowskip=3mm,
  showstringspaces=false,
  columns=flexible,
  basicstyle={\small\ttfamily},
  numbers=none,
  numberstyle=\tiny\color{black},
  keywordstyle=\color{blue},
  commentstyle=\color{dkgreen},
  stringstyle=\color{mauve},
  breaklines=true,
  breakatwhitespace=true,
  tabsize=3
}
\begin{lstlisting}
****** HIGH ALERT: GOOSE Cyber-Attack Detected! Switching to CIED ******
\end{lstlisting}

Fig.~\ref{fig:Wireshark_new} clearly depicts the Wireshark capture of GOOSE packets by the CIED when there is GOOSE cyber-attack, that causes differential PIED to get compromised and following that, CIED takes over its differential overcurrent protection function.

\paragraph{Simulation Setup and Results}

Similar to the earlier case, when there are no cyber-attacks on the digital substation, the outcome of the optimal objective value is 0, signifying that the CIED is not enabled by the SCS. However, when there is a GOOSE attack, optimal objective value becomes 21, as D = [0,1], E = 1, and F = [0,1]. This signifies that PIED 2 (i.e., differential PIED) is attacked and disabled. Hence, CIED is enabled to take over, and flow rules are updated accordingly.

Under normal operating conditions, even in the absence of cyber-attacks, various faults such as those involving line feeders and transformers can still occur. In these scenarios, the proposed SCS framework plays a vital role in ensuring the continuity of substation operations. When a fault is detected, the associated PIED is disabled as part of the SCS, and the CIED is activated.

The CIED, now fully operational, takes over the critical protection functions previously handled by the PIEDs. These functions include monitoring and responding to anomalies or faults within the substation network. Upon detecting a line feeder or transformer fault, the CIED executes the necessary protective actions, such as tripping the associated CBs. This prompt response helps isolate the faulted section, preventing damage and ensuring safety.

By leveraging the advanced capabilities of the SCS framework, which includes dynamic reconfiguration and real-time monitoring, the system ensures that any operational disruptions are minimized. This seamless transition to CIED maintains the stability and reliability of the digital substation, ensuring that essential operations continue without interruption.

Thus, the integration of the SCS framework provides a robust solution that mitigates the impact of cyber-attacks and ensures that normal operational faults are managed effectively, preserving the continuity and reliability of substation operations.

\section{Conclusion and Future Work}
The proposed SDN-based defense strategy, SCS, utilizes OpenFlow rules to prevent cyber-attacks on IEC 61850-based SV and GOOSE messages. The APC's centralized and programmable control plane allows for dynamic and efficient network management, enhancing flexibility and scalability in security policy implementation. The SCS IDS detects malicious SV and GOOSE messages, locates compromised devices, and dynamically reconfigures the network. 
The CIED concept offers redundancy by transitioning protection functions from compromised physical devices to CIEDs, maintaining operational integrity during cyber-attacks. Evaluation using a HIL testbed confirms the framework’s efficacy in maintaining substation functionality. Future work will explore algorithms for packet dropping, re-routing, and dynamic traffic redirection for real-time threat mitigation and anomaly detection. Continuous optimization of the SCS IDS architecture, incorporating machine learning or AI-based mechanisms, will enhance adaptability to evolving cyber threats.

\section{Acknowledgment}
This research was partially funded by the Director of Cybersecurity, Energy Security, and Emergency Response, specifically through the Cybersecurity for Energy Delivery Systems program of the U.S. Department of Energy under contract DE-CR0000021. The views, findings, conclusions, or recommendations presented in this material are solely those of the authors and do not necessarily represent those of the funding sponsors.

\bibliographystyle{IEEEtran}

\end{document}